# A Sensorless Control System for an Implantable Heart Pump using a Real-time Deep Convolutional Neural Network

Masoud Fetanat*, Michael Stevens, Christopher Hayward, Nigel H. Lovell, *Fellow, IEEE*

**Abstract**— Left ventricular assist devices (LVADs) are mechanical pumps, which can be used to support heart failure (HF) patients as bridge to transplant and destination therapy. To automatically adjust the LVAD speed, a physiological control system needs to be designed to respond to variations of patient hemodynamics across a variety of clinical scenarios. These control systems require pressure feedback signals from the cardiovascular system. However, there are no suitable long-term implantable sensors available. In this study, a novel real-time deep convolutional neural network (CNN) for estimation of preload based on the LVAD flow was proposed. A new sensorless adaptive physiological control system for an LVAD pump was developed using the full dynamic form of model free adaptive control (FFDL-MFAC) and the proposed preload estimator to maintain the patient conditions in safe physiological ranges. The CNN model for preload estimation was trained and evaluated through 10-fold cross validation on 100 different patient conditions and the proposed sensorless control system was assessed on a new testing set of 30 different patient conditions across six different patient scenarios. The proposed preload estimator was extremely accurate with a correlation coefficient of 0.97, root mean squared error of 0.84 mmHg, reproducibility coefficient of 1.56 mmHg, coefficient of variation of 14.44 %, and bias of 0.29 mmHg for the testing dataset. The results also indicate that the proposed sensorless physiological controller works similarly to the preload-based physiological control system for LVAD using measured preload to prevent ventricular suction and pulmonary congestion. This study shows that the LVADs can respond appropriately to changing patient states and physiological demands without the need for additional pressure or flow measurements.

*Index Terms*— Convolutional neural network, left ventricular assist devices, deep learning, model free adaptive control, physiological control, sensorless control.

*Masoud Fetanat (correspondence e-mail:m.fetanat@ieee.org), Michael Stevens and Nigel Lovell are with the Graduate School of Biomedical Engineering, UNSW, Sydney, Australia. Christopher Hayward is with Cardiology Department, St Vincent's Hospital, Sydney, NSW, Australia; Victor Chang Cardiac Research Institute, Sydney, NSW, Australia; and School of Medicine, UNSW, Sydney, NSW, Australia.

## I. INTRODUCTION

HEART failure (HF) affects more than 100,000 adults, 6.5 million and 23 million people in Australia, USA and worldwide, respectively [1]–[3]. HF prevalence in the USA is expected to increase by 46% from 2012 to 2030, lead to more than 8 million people diagnosed with HF [2]. Although heart transplantation is the gold standard treatment for end-stage HF patients, due to the lack of donor hearts, only 113 heart transplantation were performed in Australia in 2019 [4]. Mechanical circulatory support (MCS) is a standard clinical therapy for advanced HF patients, achieved by implantation of a mechanical pump in HF patients. Left ventricular assist devices (LVADs) are a type of MCS for a failing left ventricle by pumping blood from the left ventricle to the aorta. These devices are used during bridge to transplant, bridge to recovery and destination therapy [5].

Currently, clinicians set the LVAD speed in a constant mode, which can lead to hazardous events such as insufficient perfusion, ventricular suction (ventricular collapse caused by low pressure in the ventricle) or pulmonary congestion (excess fluid in the lungs due to high pressure in the ventricle). While ventricular suction can lead to hemolysis, myocardial damage, right ventricular dysfunction, ventricular thrombus and subsequent stroke, pulmonary congestion can result in pulmonary edema and shortness of breath [5]–[7]. A physiological control system, which automatically adjusts pump speed according to hemodynamic variations or pump variables, can mitigate these hazardous events. Physiological control systems can be designed to follow different objectives such as constant differential pressure [8], constant pump flow [9], constant pump inlet pressure [10] or a Starling-like controller which sets target flow rate based on the preload (left ventricular end-diastolic pressure) [11]–[15]. Whilst many physiological control systems have been proposed, there are several issues associated with these approaches.

First, most of these physiological control systems have only been evaluated using a single simulated patient or condition [13]–[18], which does not represent the wide clinical range of interpatient and intrapatient variations in cardiovascular system (CVS) dynamics. Furthermore, in most studies of physiological control systems for LVADs [13]–[15], [18], a proportional-integral-derivative (PID) controller was employed, which was tuned for a specific patient and condition, and control performance cannot be guaranteed across different patients and conditions, which may lead to hazardous events [19]. In recent studies, adaptive control systems such as artificial neural network (ANN) control [17] and fuzzy logic control (FLC) [20], which can automatically adjust their parameters according to feedback from the controlled system, have been used in the control of LVADs. However, these controllers need large training data and precise determination of rules for different patient conditions, respectively. Fetanat *et al.* proposed a preload-based physiological control for an LVAD using model free adaptive control (MFAC) to maintain the preload between 3 to 15 mmHg, which was validated on 100 different patient conditions in which control performance can be guaranteed across different patients and conditions [19], [21]. However, the compact linearization form of MFAC was employed to control the LVAD speed, which cannot identify the full dynamics of the controlled system and therefore can lead to inaccurate control. The full dynamic linearization form of model free adaptive control (FFDL-MFAC) could be used to identify the dynamic characteristics of the CVS and LVAD using the memories of the past inputs and outputs of the CVS and LVAD without knowledge of the mathematical model or dynamic information of the controlled system. FFDL-MFAC also does not require any training process or rules like ANNs and fuzzy controllers.

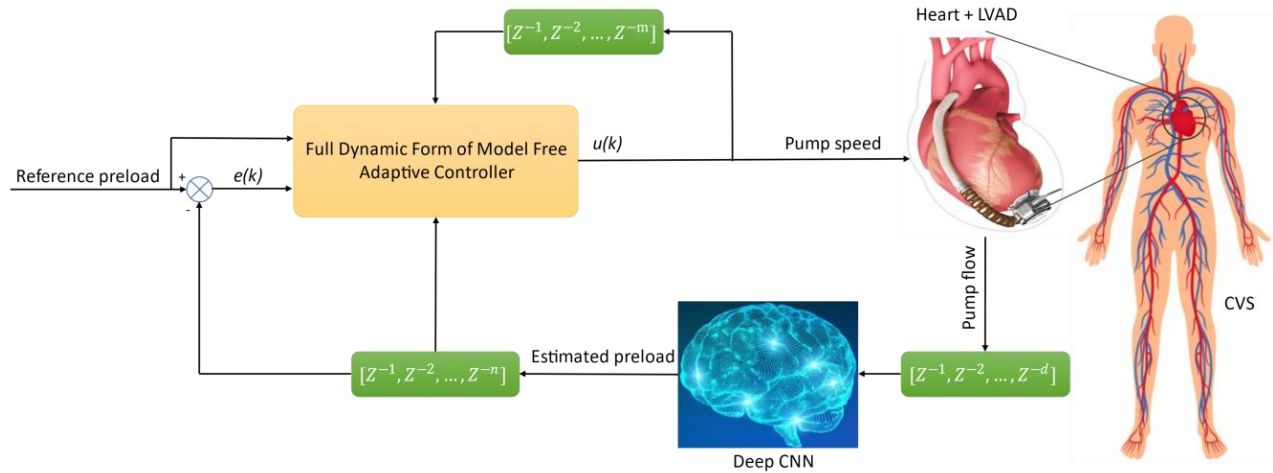

Fig. 1. Block diagram of the proposed sensorless physiological control system for LVAD. u(k) is the control signal (pump speed), and e(k) is the error between reference and estimated preload. m, d and n are the number of delays from pump speed, pump flow and estimated preload, respectively. CNN: convolutional neural network, CVS: cardiovascular system and LVAD: left ventricular assist device. Heart + LVAD (reproduced with permission of Medtronic).

Additionally, while most of the current physiological control systems require direct measurement of pressure [18], [19], [21], flow [22] or ventricular volume [23] via implantation of a sensor, these sensors are faced with several challenges such as drift, additional power consumption, interference from radiation, thrombus formation and device failure [5]. These challenges have limited real-world application, leading researchers to focus on estimating these measurements. Although these measurements can be estimated using machine learning (ML) algorithms, finding appropriate features as inputs for these algorithms is a difficult task. ML methods are a part of artificial intelligence (AI), which can learn a pattern from a training dataset and then accurately predict output on a new dataset [24]. ML algorithms have been widely used in classification and regression problems for cardiovascular medicine [25]. Deep learning (DL) is a new type of ML, which is usually made by several layers of a convolutional neural network (CNN), pooling and a multi-layer perceptron (MLP) [25]. Applying deep CNNs for estimation of preload can automatically extract suitable features from the input signals, using pump flow in real-time mode across the range of interpatient and intrapatient CVS variations.

The aim of this study is to design a sensorless adaptive physiological control system for an implantable LVAD using deep CNNs and FFDL-MFAC to provide consistent control performance across interpatient and intrapatient variations without needing any implantable flow or pressure sensors. The objective of the proposed control system is to prevent hazardous events such as ventricular suction and pulmonary congestion and maintain the patient conditions in the physiological range, without reliance on an implantable sensor. To our knowledge, this is the first study, where a CNN was employed for the purpose of estimation combined with a controller in a real-time mode.

In this paper, first the numerical model of the human CVS and LVAD for the evaluation of the proposed sensorless physiological control system is introduced. Afterward, the novel real-time method for preload estimation using deep CNNs is presented. Then, the FFDL-MFAC and different patient conditions and scenarios used for evaluation of the sensorless physiological control system are described. Subsequently, the results of the proposed preload estimation and sensorless adaptive physiological control system using deep CNN and FFDL-MFAC across interpatient and intrapatient variations is presented. Finally, a discussion including merits and limitations of the study and future work is presented.

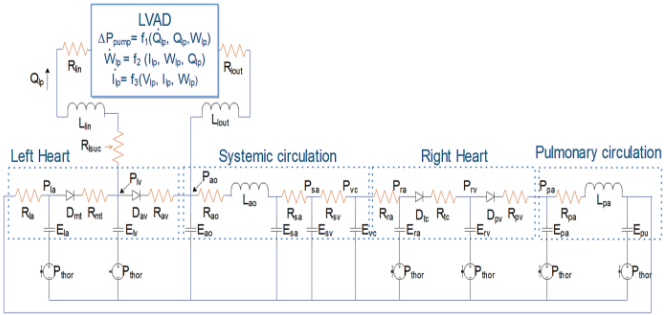

Fig. 2. Electrical equivalent circuit analogue of cardiovascular system with LVAD interaction model. P, pressures; R, resistances; E, elastances (=1/compliances); L, inertances; D, diodes. The model consists of two main components: (1) the cardiovascular model includes left heart, systemic circulation, right heart, pulmonary circulation (la, left atrium; lv, left ventricle, ao, aorta; sa, systemic peripheral vessels, including the arteries and capillaries; sv, systemic veins, including small and large veins; vc, vena cava; ra, right atrium; rv, right ventricle; pa, pulmonary peripheral vessels, including pulmonary arteries and capillaries; pu, pulmonary veins and (2) LVAD model includes Rin and Rout, inlet and outlet cannulae resistances; Lin and Lout, inlet and outlet cannulae inertances; Rlsuc, left suction resistance, Rrsuc, right suction resistance, Rband, banding resistance. The intrathoracic pressure, Pthor was assigned –4 mmHg during closed-chest simulated conditions [19].

## II. METHODOLOGY

Fig. 1 demonstrates the block diagram of the proposed estimator and controller evaluated in this study. It includes three main blocks: the model of the CVS and LVAD, deep CNN for preload estimation and FFDL-MFAC system.

### A. Model of Cardiovascular System and LVAD

In this work, the implementation of the model of cardiovascular system and LVAD were performed in Simulink (MathWorks, Natick, MA, USA). Fig. 2 shows the numerical model of the cardiovascular system used in this study, which was proposed by Lim *et al.* in 2010 [26]. The CVS model includes systemic and pulmonary circulations, left and right hearts and LVAD inlet and outlet cannulae. It was derived using first principles and validated using in-vitro tests and in-vivo animal experiments [26]–[28].

The HVAD (Medtronic, Minneapolis, MN, USA) centrifugal pump was employed as the LVAD in this study. The HVAD pump model used was derived from the model proposed by Boes *et al.* [29] including the pump and peripheral models of components connected to the pump. The speed range of the HVAD pump model was limited to 1800 and 4000 rpm, which is the clinical operating range of the device.

### B. Deep Convolutional Neural Network for Preload Estimation

Deep CNNs are hierarchical neural networks and one of the most popular deep learning methods. The structure of CNNs commonly include convolutional, pooling (subsampling), fully-connected and dropout layers [25], [30]. While in the convolutional layer, the features of input signal are automatically extracted using different filters, in the dropout layer, only essential features are maintained using subsampling operations [31]. A fully-connected layer basically works as a conventional MLP, which includes neurons connected to each other using weights, biases and activation functions. A dropout layer removes unnecessary neuron connections [31].

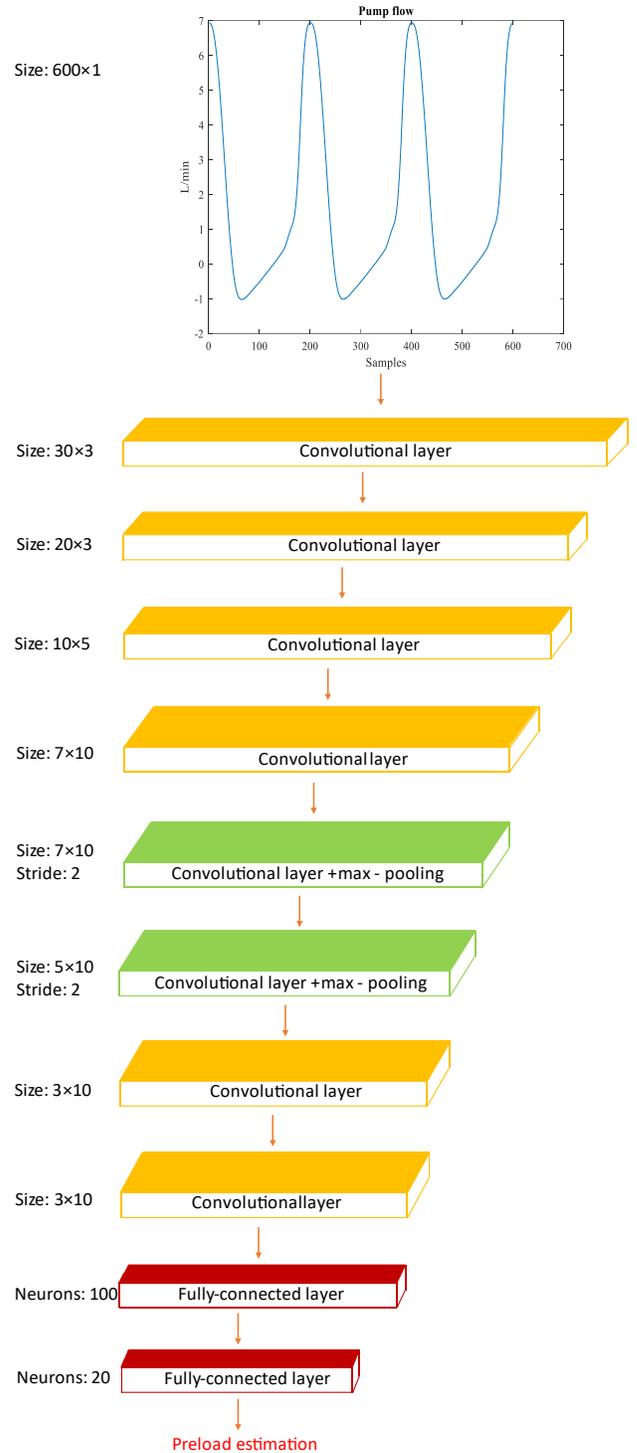

Fig. 3. Architecture of the proposed deep CNN model for estimation of preload. 600 samples from the peak of the pump flow signal in each cardiac cycle are fed to the CNN as input. Afterwards, it is followed by four convolutional layers including batch normalization layers. Subsequently, two convolutional, max-pooling layers with stride 2 and batch normalization layers are placed. Then, it is followed by two convolutional layers including batch normalization layers. Finally, fully-connected layers with 100 and 20 neurons, activation function of LeakyReLU and a dropout layer with probability of 20% are then used to estimate preload.

Fig. 3 depicts the architecture of the proposed deep CNN model for estimation of preload based on the 600 samples from the peak of the pump flow. The whole process can be described as follows: first, in each cardiac cycle the peak of pump flow is identified using a heuristic peak detector. Once the peak is detected 600 samples (three seconds) of the pump flow before the peak is fed to the deep CNN model. The number of samples fed to deep CNN was optimized to achieve minimum root mean square error (RMSE) between actual and estimated preload through 10-fold cross validation. It is then followed by four convolutional layers with the filter size of 30×3, 20×3, 10×5 and 7×10. Each convolutional layer includes one batch normalization layer. Afterward, two convolutional and max-pooling layers with the filter size of 7×10 and 5×10 and stride length of 2 are placed. Subsequently, it is followed by two convolutional layers with the filter size of 3×10. Finally, fully-connected layers with 100 and 20 neurons, an activation function of LeakyReLU and a dropout layer with probability of 20% are employed to estimate preload. All these parameters were empirically adjusted to minimize RMSE between actual and estimated preload through 10-fold cross validation.

The proposed deep CNN method for preload estimation was evaluated using the same method previously used by our group [19]. Briefly, the estimation method was applied in six different patient scenarios; rising and falling pulmonary vascular resistance (PVR), rising and falling systemic vascular resistance (SVR), rest to exercise and passive postural changes. Each scenario was simulated using 100 different patient conditions and four speeds of the HVAD pump (2000, 2600, 3000 and 3400), which generated preloads between -1 to 25 mmHg. SVR and PVR variations were performed by changes in systemic arterial resistance ($R_{sa}$) and pulmonary arterial resistance ($R_{pa}$) since the arterial resistances contribute a major part of total vascular resistance in the CVS model. Evaluation of the deep CNN model for estimation of preload was performed using 100 different LVAD patients for each of the six scenarios.

To compare the difference between measured and estimated preload, correlation coefficient (R), RMSE, reproducibility coefficient (RPC), coefficient of variation (CV) and bias were used in terms of Bland-Altman analysis [32]. RMSE can be defined by (1):

$$RMSE = \sqrt{\frac{1}{n}\sum_{i=1}^{n}(actual\ preload_i - estimated\ preload_i)^2} \quad (1)$$

where $actual\ preload_i$ and $estimated\ preload_i$ are measured and estimated preload at time instant of $i$ and $n$ is the number of total samples in each simulation. The sampling rate of measured variables is 200 Hz.

RPC or limits of agreement and CV are defined as follows respectively: $RPC = 1.96 * standard\ deviation$ and $CV = standard\ deviation\ over\ mean\ values\ of\ estimated\ and$ $measured\ preload$. Bias presents the mean of difference between estimated and actual preload. The less RMSE, RPC and bias, the more accurate preload estimation. CV can be referred as a measure of precision. Although the acceptable range of CV is different in each research field, CV of 20% or less is generally assumed as an acceptable range [33]. However, the lower value of the CV, the more precise the estimate [33].

The proposed deep CNN was trained using the Adam algorithm [34] with 1000 iterations, a batch size of 1102 and the He algorithm [35] for weights initialization.

TABLE I
NORMAL VALUES AND DESCRIPTION OF CARDIOVASCULAR MODEL PARAMETERS FOR INTERPATIENT AND INTRAPATIENT SIMULATIONS

| No | Parameter (unit) | Description | Normal Value |
|---|---|---|---|
| 1 | Eeslvf (mmHg.mL$^{-1}$) | LV end systolic elastance | 3.54 |
| 2 | Eesrvf (mmHg.mL$^{-1}$) | RV end systolic elastance | 1.75 |
| 3 | Eao (mmHg.mL$^{-1}$) | Aortic elastance | 1.04 |
| 4 | Eesla (mmHg.mL$^{-1}$) | LA end systolic elastance | 0.2 |
| 5 | Eesra (mmHg.mL$^{-1}$) | RA end systolic elastance | 0.2 |
| 6 | Epa (mmHg.mL$^{-1}$) | Pulmonary arterial elastance | 0.15 |
| 7 | Epu (mmHg.mL$^{-1}$) | Pulmonary vein elastance | 0.04 |
| 8 | Esa (mmHg.mL$^{-1}$) | Systemic arterial elastance | 0.37 |
| 9 | Esv (mmHg.mL$^{-1}$) | Systemic vein elastance | 0.013 |
| 10 | Evc (mmHg.mL$^{-1}$) | Vena cava elastance | 0.03 |
| 11 | Rao (mmHg.s.mL$^{-1}$) | Aortic resistance | 0.2 |
| 12 | Rra (mmHg.s.mL$^{-1}$) | Right atrium resistance | 0.012 |
| 13 | Rpv (mmHg.s.mL$^{-1}$) | Pulmonary valve resistance | 0.02 |
| 14 | Rsv (mmHg.s.mL$^{-1}$) | Systemic venous resistance | 0.12 |
| 15 | Tc (s) | Heart rate coefficient | 1 |
| 16 | Tsys0 (s) | Maximum systolic heart period | 0.5 |
| 17 | V0la (mL) | LA end diastolic volume at zero pressure | 20 |
| 18 | V0lvf (mL) | LV end diastolic volume at zero pressure | 40 |
| 19 | V0ra (mL) | RA end diastolic volume at zero pressure | 20 |
| 20 | V0rvf (mL) | RV end diastolic volume at zero pressure | 50 |
| 21 | Vdla (mL) | LA end systolic volume at zero pressure | 10 |
| 22 | Vdlvf (mL) | LV end systolic volume at zero pressure | 16.77 |
| 23 | Vdra (mL) | RA end systolic volume at zero pressure | 10 |
| 24 | Vdrvf (mL) | RV end systolic volume at zero pressure | 40 |
| 25 | Rmt (mmHg.s.mL$^{-1}$) | Mitral valve resistance | 0.01 |
| 26 | Rav (mmHg.s.mL$^{-1}$) | Aortic valve resistance | 0.02 |
| 27 | Vuao (mL) | Aortic unstressed volume | 230.88 |
| 28 | Vupa (mL) | Pulmonary arterial unstressed volume | 91.67 |
| 29 | Vupu (mL) | Pulmonary vein unstressed volume | 132.39 |
| 30 | Vusa (mL) | Systemic arterial unstressed volume | 231.04 |
| 31 | Vusv (mL) | Systemic vein unstressed volume | 1976.1 |
| 32 | Vuvc (mL) | Vena cava unstressed volume | 136.17 |
| 33 | P0la (mmHg) | LA end diastolic stiffness scaling term | 0.5 |
| 34 | P0lvf (mmHg) | LV end diastolic stiffness scaling term | 0.98 |
| 35 | P0ra (mmHg) | RA end diastolic stiffness scaling term | 0.5 |
| 36 | P0rvf (mmHg) | RV end diastolic stiffness scaling term | 0.91 |
| 37 | Vtotal (mL) | Total blood volume | 5200 |
| 38 | λla (mL$^{-1}$) | LA end diastolic stiffness coefficient | 0.025 |
| 39 | λlvf (mL$^{-1}$) | LV end diastolic stiffness coefficient | 0.028 |
| 40 | λra (mL$^{-1}$) | RA end diastolic stiffness coefficient | 0.025 |
| 41 | λrvf (mL$^{-1}$) | RV end diastolic stiffness coefficient | 0.028 |
| 42 | Lao (mmHg.s$^2$.ml$^{-2}$) | Aortic inertance | 0.0001 |
| 43 | Lpa (mmHg.s$^2$.ml$^{-2}$) | Pulmonary arterial inertance | 7.70e-05 |

## C. Full Dynamic Form of Model Free Adaptive Control (FFDL-MFAC)

MFAC is a type of data-driven control (DDC) system in which the controller is designed by input and output data achieved from the controlled system in an online mode instead of using a mathematical model or system structure of the controlled system. Our group has previously shown that a control system using an MFAC produces more consistent control performance across a range of different patient scenarios than PID control [19]. However, one of the limitations of using MFAC is that the control input is determined only by single input and output samples. This can be overcome by a full form dynamic linearization technique (FFDL). FFDL-MFAC is a type a MFAC which employs FFDL and pseudo gradient (PG) to capture full dynamic behavior of the controlled system using the memories of the past inputs and outputs of the controlled system [36], [37]. The fundamental operation of FFDL-MFAC is described in [36], [37], with a brief description as follows.

An unknown nonaffine single input single output (SISO) nonlinear discrete-time system can be presented as follows:

$$y(k+1) = f(y(k), y(k-1), \ldots, y(k-n_y), u(k), u(k-1), \ldots, u(k-n_u)) \quad (2)$$

where $y(k) \epsilon R$ and $u(k) \epsilon R$ are the system output and control input at time instant $k$ respectively, $n_y$ and $n_u$ are unknown orders of output and input, and $f(.): R^{n_u+n_y+2} \to R$ is an unknown nonlinear function.

The full dynamic linearization form of nonlinear system (2) can be defined by (3):

$$y(k+1) = y(k) + \varphi_{f, L_u, L_y}^T(k) \Delta H_{L_y, L_u}(k)$$
$$\varphi_{f, L_u, L_y}(k) = [\varphi_1(k), \ldots, \varphi_{L_y-1}(k), \varphi_{L_y}(k), \ldots, \varphi_{L_u+L_y}(k)]^T$$
$$H_{L_y, L_u}(k) = [y(k), \ldots, y(k-L_y+1), u(k), \ldots, u(k-L_u+1)]$$
$$\Delta H_{L_y, L_u}(k) = H_{L_y, L_u}(k) - H_{L_y, L_u}(k-1) \quad (3)$$

where $\varphi_{f, L_u, L_y}(k)$ is called PG and $H_{L_y, L_u}(k)$ is a vector, including of all control input and system output signals within a moving time window.

For finding the control input $u(k)$, the following cost function is minimized to have less error (difference between the desired and system output) and changes of two consecutive control inputs:

$$J(u(k)) = \|y_d(k+1) - y(k+1)\|^2 + \lambda \|u(k) - u(k-1)\|^2 \quad (4)$$

where $y_d(k+1)$ and $\lambda > 0$ are desired output and a weighting factor respectively.

By substituting (3) into (4) and differentiating (4) with respect to control signal $u(k)$ equals to zero, the control input is derived as follows:

TABLE II
SIX DIFFERENT PATIENT SCENARIOS AND THEIR SIMULATION CONDITIONS

| Patient scenarios | Simulation conditions | Reference |
|---|---|---|
| Increasing $R_{pa}$ changes | 100 to 500 dyne.s.cm$^{-5}$ | [19] |
| Decreasing $R_{pa}$ changes | 100 to 40 dyne.s.cm$^{-5}$ | [19] |
| Increasing $R_{sa}$ changes | 1300 to 2600 dyne.s.cm$^{-5}$ | [19] |
| Decreasing $R_{sa}$ changes | 1300 to 600 dyne.s.cm$^{-5}$ | [19] |
| Transition from rest to exercise | 1. Increasing heart rate from 60 to 80 bpm<br>2. Decreasing $R_{pa}$ from 100 to 40 dyne.s.cm$^{-5}$<br>3. Decreasing $R_{sa}$ from 1300 to 670 dyne.s.cm$^{-5}$<br>4. Adding 500 mL fluid from systemic vein into the right atrium | [19] |
| Passive postural change | Removing 300 mL fluid from the volume of systemic arteries | [19] |

$$u(k) = u(k-1) + \frac{\varphi_{L_y+1}(k)[\rho_{L_y+1}(y_d(k+1) - y(k))]}{\lambda + \|\varphi_{L_y+1}(k)\|^2}$$
$$- \frac{\sum_{i=1}^{L_y} \rho_i \varphi_i(k) \Delta y(k-i+1)]}{\lambda + \|\varphi_{L_y+1}(k)\|^2}$$
$$- \frac{\varphi_{L_y+1}(k) \sum_{i=L_y+2}^{L_y+L_u} \rho_i \varphi_i(k) \Delta u(k+L_y-i+1)]}{\lambda + \|\varphi_{L_y+1}(k)\|^2} \quad (5)$$

where $\rho_i \epsilon (0,1], i = 1, \ldots, L_y + L_u$ are step constants which are added to make (5) more flexible.

A suitable parameter estimation algorithm for capturing the time-varying characteristics of PG needs to be employed. The cost function for estimation of PG is defined as follows:

$$J\left(\varphi_{f, L_u, L_y}(k)\right) = \left|\Delta y(k) - \varphi_{f, L_u, L_y}^T(k) \Delta H_{L_y, L_u}(k-1)\right|^2$$
$$+ \mu \left\|\varphi_{f, L_u, L_y}(k) - \hat{\varphi}_{f, L_u, L_y}(k)\right\|^2 \quad (6)$$

where $\mu > 0$ is a weighing factor and $\hat{\varphi}_{f, L_u, L_y}(k) \epsilon R^{L_u+L_y}$ is the estimated value of $\varphi_{f, L_u, L_y}(k)$.

By minimizing the objective function (6) with respect to $\hat{\varphi}(k)$, the PG estimation algorithm is found as follows:

$$\hat{\varphi}_{f, L_u, L_y}(k) = \hat{\varphi}_{f, L_u, L_y}(k-1)$$
$$+ \frac{\eta \Delta H_{L_y, L_u}(k-1)(y(k) - y(k-1))}{\mu + \|\Delta H_{L_y, L_u}(k-1)\|^2}$$
$$- \frac{\eta \Delta H_{L_y, L_u}(k-1) \hat{\varphi}_{f, L_u, L_y}^T(k-1) \Delta H_{L_y, L_u}(k-1)}{\mu + \|\Delta H_{L_y, L_u}(k-1)\|^2}$$
$$\hat{\varphi}_{f, L_u, L_y}(k) = [\hat{\varphi}_1(k), \ldots, \hat{\varphi}_{L_y-1}(k), \hat{\varphi}_{L_y}(k), \ldots, \hat{\varphi}_{L_u+L_y}(k)]^T \quad (7)$$

$$\hat{\varphi}_{f, L_u, L_y}(k) = \hat{\varphi}_{f, L_u, L_y}(1), if \ \|\hat{\varphi}_{f, L_u, L_y}(k)\| \leq \varepsilon$$
$$or \ \|\Delta H_{L_y, L_u}(k-1)\| \leq \varepsilon$$
$$or \ sign(\hat{\varphi}_{f, L_u, L_y}(k)) \neq sign(\hat{\varphi}_{f, L_u, L_y}(1)) \quad (8)$$

where $\eta \in (0,2]$ is a step constant which is added to make (7) flexible, $\hat{\varphi}_{f,L_u,L_y}(1)$ is the initial estimated value of $\hat{\varphi}_{f,L_u,L_y}(k)$ and $\varepsilon$ is a very small positive constant. The reset algorithm (8) is employed strengthen the ability to estimate the time-varying parameter by the parameter estimation algorithm. The FFDL-MFAC works on the following four assumptions:

Assumption 1: The partial derivative of $f(.)$ in (2) with respect to all the variables are continuous.

Assumption 2: (2) satisfies the following generalized Lipschitz condition:

$$|y(k_1+1) - y(k_2+1)| \leq b \left\| H_{L_y,L_u}(k_1) - H_{L_y,L_u}(k_1) \right\| \quad (9)$$

where $H_{L_y,L_u}(k) \in R^{L_u+L_y}$. This assumption means that the output changes are bounded by the variations of control input $(u)$ and system output $(y)$ at $k$ and past time instances.

Assumption 3: There exists a bounded control input $u^*(k)$ such that the system output reaches the desired output signal $y_d(k)$, which means the system is controllable.

Assumption 4: The sign of $\varphi_{L_y+1}(k)$ is assumed to be unchanged, i.e., $\varphi_{L_y+1}(k) > \varepsilon > 0$ or $\varphi_{L_y+1}(k) < -\varepsilon < 0$, where $\varepsilon$ is a small positive constant. This assumption means that the system output will not decrease if the difference control input $\Delta u$ increases or vice versa.

If the nonlinear system (2) satisfying assumption 1-4 is controlled by FFDL-MFAC method through (5), (7) and (8) with a constant desired signal; $y_d(k+1) = y_d = const$, then there exists a $\lambda_{min}$ such that $\lambda > \lambda_{min}$ and the closed loop system guarantees:
1. $u(k)$ and $y(k)$ are bounded for all $k$ and the closed loop system is bounded input, bounded output (BIBO) stable.
2. The system output tracking error asymptotically converges to zero; $\lim_{k \to \infty} |y_d - y(k+1)| = 0$
3. The closed-loop system is internally stable.

The model of CVS used in this study stratify all the four above assumptions. In addition, our group has previously shown that the CVS model satisfies the four similar assumptions [19].

### D. Different Patient Conditions and Scenarios

In order to simulate interpatient and intrapatient variations, all of the CVS model parameters from TABLE I were varied from -20% to +20% of their nominal values to create 100 different patient conditions for training and validation of deep CNN for preload estimation. In the same manner, 30 new patient conditions were generated for testing FFDL-MFAC on unseen simulated patients. The FFDL-MFAC combined with measured or estimated preload using the deep CNN was evaluated in each patient for six different scenarios.

These six scenarios were simulated by variation of different vascular resistances, volumes, and heart rate. These scenarios were rapid $R_{pa}$ changes (rising and falling), rapid $R_{sa}$ changes (rising and falling), transition from rest to exercise and passive postural change experiments [19]. TABLE II shows the simulation conditions for each scenario. The CVS parameters in transition from rest to exercise and passive postural change scenarios were changed using a first order system with a time constant of 10 s, which simulates the response of the native CVS. The other scenarios were simulated using a step-change in parameters [19]. These rapid time-domain responses were chosen because if the proposed methodology can respond properly to these extreme scenarios, it can also respond to slower and milder changes appropriately [13].

### E. Evaluation of Sensorless Control

Evaluation of the FFDL-MFAC system for controlling of the HVAD pump speed was performed across interpatient and intrapatient variations using 30 different LVAD patients for each of the six scenarios. The 43 CVS model parameters were randomly varied between -20% to 20% from their nominal values at the beginning of the experiments to create each new patient. Importantly, these were 30 "unseen" simulated patients, in which the preload estimator was not trained using any of these 30 patients.

In each simulation for each scenario and each patient, an FFDL-MFAC system was implemented to maintain preload (measured or estimated) at a target constant value based on the following protocol. In the first 25 seconds, the pump was off. In the next 25 seconds, the heart pump works in a constant speed mode, in which the speed is tuned to have cardiac output and mean aortic pressure (MAoP) in the range of 4 to 6 l/min and 70 to 90 mmHg, respectively. This process is similar to the actions that clinicians perform to adjust the pump speed for LVAD patients. The 25 second period allows the CVS system to reach a steady state. The desired (target) preload is then set as the preload during this steady state period. At time 50 s, the FFDL-MFAC was then activated. At time 70 s, one of the six patient scenarios was simulated and finally at time 120 s. the simulations ended. The FFDL-MFAC parameters were chosen empirically as $L_u = 3$, $L_y = 3$, $\mu = 1$, $\lambda = 1$, $\eta = 0.3, \varepsilon = 10^{-4}$, $\varphi(1) = [1,1,1,1,1,1]^T$ and $\rho = [0.05, 0.05, 0.05, 0.05, 0.05, 0.05]$. Ventricular suction and pulmonary congestion were defined based on the preload below 0 and above 20 mmHg, respectively. The occurrence of these two important hazardous events was used to assess preload-based FFDL-MFAC for measured and estimated preload. The experiments were performed on a computer with 16 GB of RAM, CPU of Core i7-8700K 3.70 GHz, GPU of GTX 1650 SUPER Windforce 4GB and MATLAB software of 2020b.

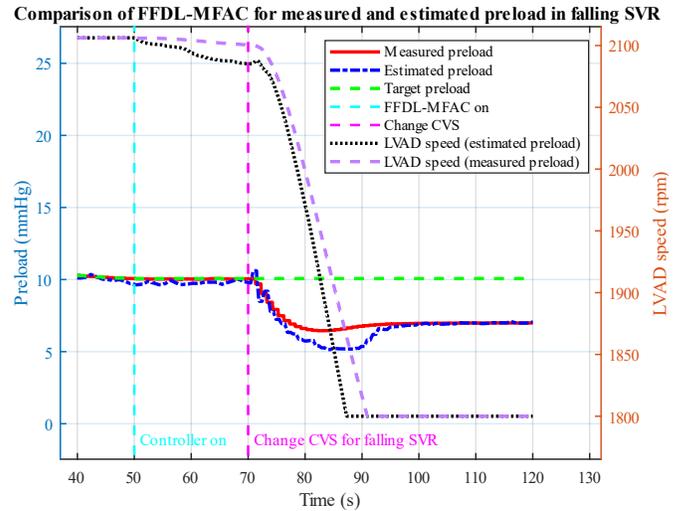

Fig. 4. A sample simulation for comparison of FFDL-MFAC for measured and estimated preload in falling SVR (patient 23).

TABLE III
PERFORMANCE OF THE PROPOSED DEEP CNN ON THE ESTIMATION OF PRELOAD FOR EACH VALIDATION FOLD THROUGH 10-FOLD CROSS-VALIDATION ON 100 PATIENTS

| Fold | Correlation coefficient | RMSE (mmHg) | RPC (mmHg) | CV (%) | Bias (mmHg) |
|---|---|---|---|---|---|
| 1 | 0.98 | 1.00 | 1.97 | 17.17 | -0.09 |
| 2 | 0.98 | 0.83 | 1.64 | 14.37 | -0.02 |
| 3 | 0.99 | 0.85 | 1.67 | 14.71 | 0.06 |
| 4 | 0.99 | 0.70 | 1.39 | 12.24 | 0.05 |
| 5 | 0.99 | 0.80 | 1.57 | 13.85 | 0.03 |
| 6 | 0.99 | 0.80 | 1.57 | 13.78 | -0.01 |
| 7 | 0.99 | 0.88 | 1.73 | 15.29 | 0.07 |
| 8 | 0.98 | 1.13 | 2.21 | 19.57 | 0.07 |
| 9 | 0.98 | 0.95 | 1.87 | 16.44 | 0.02 |
| 10 | 0.99 | 0.89 | 1.76 | 15.52 | 0.04 |
| **Average** | **0.99** | **0.88** | **1.73** | **15.29** | **0.02** |

TABLE IV
PERFORMANCE OF THE PROPOSED DEEP CNN COMBINED WITH FFDL-MFAC ON THE ESTIMATION OF PRELOAD FOR EACH SCENARIO ON NEW 30 PATIENTS

| Scenario | Correlation coefficient | RMSE (mmHg) | RPC (mmHg) | CV (%) | Bias (mmHg) |
|---|---|---|---|---|---|
| Falling $R_{pa}$ | 0.98 | 0.90 | 1.26 | 10.83 | 0.63 |
| Rising $R_{pa}$ | 0.97 | 0.72 | 1.41 | 16.87 | 0.11 |
| Falling $R_{sa}$ | 0.97 | 0.68 | 1.21 | 13.70 | 0.30 |
| Rising $R_{sa}$ | 0.97 | 0.94 | 1.78 | 12.75 | 0.27 |
| Rest to exercise | 0.94 | 1.07 | 2.09 | 18.49 | 0.13 |
| Postural change | 0.98 | 0.63 | 1.13 | 10.65 | 0.27 |
| **All scenarios** | **0.97** | **0.84** | **1.56** | **14.44** | **0.29** |

TABLE V
COMPARISON OF THE PROPOSED PHYSIOLOGICAL CONTROL SYSTEM USING FFDL-MFAC FOR MEASURED AND ESTIMATED PRELOAD ON NEW 30 PATIENTS IN SIX DIFFERENT SCENARIOS

| Methodology | Number of suctions | Number of congestions |
|---|---|---|
| FFDL-MFAC using measured preload | 0 | 0 |
| FFDL-MFAC using estimated preload | 0 | 0 |

In this study, 10-fold cross validation was used to validate the results of preload estimation using the proposed deep CNN. In 10-fold cross validation, the original data is randomly split into 10 equal sections. One out of 10 sections is used as the validation data and the remaining nine sections are used as the training data. This process is then repeated 10 times to each of 10 sections as the validation data. The advantage of 10-fold cross validation is that each section is used for validation once and all 10 sections are used for both training and validation, which leads to more robust results [38].

## III. RESULTS

### A. Preload Estimation Using Deep Convolutional Neural Network

TABLE III shows the result of the proposed deep CNN model for estimation of preload through 10-fold cross validation for 100 different patient conditions and four speed (2000, 2600, 3000 and 3400 rpm) in six different patient scenarios (rising and falling $R_{pa}$, rising and falling $R_{sa}$, rest to exercise and postural change). The results show that the proposed method has a small RMSE (0.88 mmHg) for all simulations (2400 simulations). The average values of the correlation coefficient, RPC, CV and bias for all the folds are 0.99, 1.73 mmHg, 15.29 % and 0.02 mmHg, respectively.

### B. Evaluation of Sensorless Control Using Deep Convolutional Neural Network and FFDL-MFAC

Fig. 4 demonstrates a sample simulation for the combined FFDL-MFAC and deep CNN model for estimation of preload and its comparison with the FFDL-MFAC and measured preload for adjusting the LVAD speed. The simulation was performed for falling SVR from 1300 to 600 dyne.s.cm$^{-5}$ in patient 23. As can be seen from the figure, there is only a small difference between the estimated and measured preload before and after the CVS change. In addition, the LVAD speed curve for both controllers are very similar. This indicates that the estimator is working accurately and that the control performance is minimally effected by the use of an estimator.

TABLE IV shows the performance of the proposed deep CNN combined with FFDL-MFAC on the estimation of the preload for six different scenarios on new 30 patients.

In transition from rest to exercise scenario, the results show a correlation coefficient of 0.94, RMSE of 1.07 mmHg, RPC of 2.09 mmHg, CV of 18.49 % and bias of 0.13 mmHg. Rising $R_{pa}$ scenario has the least bias, while falling $R_{pa}$ has the most bias. The highest CV, RPC and RMSE were produced by the rest to exercise scenario and the postural change scenario produced the least RMSE and RPC.

Finally, the results for all scenarios show a correlation coefficient of 0.97, RMSE of 0.84 mmHg, RPC of 1.56 mmHg, CV of 14.44 % and bias of 0.29 mmHg.

Futhermore, none of 30 simulations in all six scenarios (180 simulations) led to ventricular suction or pulmonary congestion, as shown in TABLE V.

## IV. DISCUSSION

In this study, we proposed a real-time deep CNN model for estimation of preload based on the pump flow, which was trained and validated across a range of 2400 randomized patient conditions. In addition, a sensorless physiological control system using FFDL-MFAC and deep CNN for preload estimation was proposed for adjusting LVAD speed to respond to interpatient intrapatient variations. The results of the preload estimator indicate that the proposed estimator can predict preload with a low amount of error. The results for the sensorless physiological control system showed that there was not any pulmonary congestion or ventricular suction when the control system was subject to 30 unseen patient conditions across six different patient scenarios. Our proposed method can work on both adults and pediatric patients, as the mean aortic valve flow and pump flow for the 30 unseen patient datasets are in the range of [0, 4.69] and [0.83, 5.87] L/min (when the proposed controller is on), respectively. The novelty of this work is that it is the first time (to the knowledge of the authors) a study has combined an accurate preload estimator with a robust control algorithm and tested it across a wide range of patient scenarios.

The results from TABLE IV demonstrate that the proposed deep CNN can estimate preload accurately based on the 600 samples (from the peak) of the pump flow on unseen patient dataset over six common patient scearios. The advantage of deep CNN over conventional machine learning algorithms like MLPs and support vector machines (SVM) is that features from the input signal are automatically extracted by applying different sizes of convolutional filters and functions. Manually finding suitable features from the input signals is a difficult and time-consuming task, as it relies on both deep understanding of the morphology of the HVAD signals and an iterative approach to identifying useful features for the estimation model.

Other researchers have presented pressure estimation methods in the literature, but none have been as accurate nor as robustly assessed as that presented in this paper. An adaptive physiological controller for rotary blood pumps using pump intrinsic variables and a linear state-space model of CVS was proposed by Wu [39]. The pump flow and aortic pressure were estimated via pump intrinsic variables. However, the pump flow derived from pump intrinsic variables was assumed to be the total flow of the CVS [39], which makes the simplified assumption that the aortic valve is always closed. Moreover, the model linearization cannot be used to simulate different patient conditions which can result in inaccuracy of simulation and inconsistent physiological control performance. Furthermore, there was no strategy to avoid the possible suction and congestion events [39].

A sensorless control for continuous flow LVAD was proposed by Meki *et al.* to maintain physiological perfusion and avoid ventricular suction using pump speed [40]. The proposed controller was used to maintain the differential pump speed setpoint of 800 rpm. However, unlike our proposed solution, the simulations were only performed on one patient with different scenarios and therefore it is not clear how the proposed methodology can respond to different patient conditions and how the differential pump speed setpoint should be set for different patients.

The slope of the pump flow over the ventricular filling phase was used to estimate pulmonary arterial wedge pressure (PAWP) as a surrogate of preload using linear regressions [41]–[43]. However, in these studies, estimation of PAWP was according to distinguish PAWP to two classes of high PAWP (PAWP $\geq$ 18 or PAWP $\geq$ 30) and low PAWP (PAWP $\leq$ 18), which make them impossible to be used in real-time physiological control of LVADs. HeartPod as an implantable pressure sensor for measuring left atrial pressure resulted in an error of 0.8 ± 4.0 mmHg, which can lead to 4.7 mmHg drift after three months [44]. However, in this study, preload can be accurately estimated with error of about 0.9 mmHg in the worst case. An inaccurate estimator can result in poorer control performance. Furthermore, we have shown that even with this error, there is little difference in control performance between using our estimator and a sensor.

The use of the proposed FFDL-MFAC preload-based control system in this paper showed excellent performance across a wide range of unseen patient conditions. FFDL-MFAC can automatically adjust the pump speed in an adaptive mode using the memories of the past inputs and output from the CVS and LVAD in real-time. The adaptive structure of the FFDL-MFAC can respond to the interpatient and intrapatients variations and unpredicted changes on hemodynamics.

The mathematical proof for FFDL-MFAC shows that the closed-loop system is internally and BIBO stable and the tracking error between desired and actual output asymptotically converges to zero. Another advantage of FFDL-MFAC over the conventional PID controller used for physiological control of LVAD is that no tuning and optimization in needed for the parameters of the controller, which can facilitate working with physiological control systems for clinicians without any knowledge of control engineering.

In healthy hearts, cardiac output (CO) varies based on the preload via Frank-Starling (FS) mechanism. Therefore, preload is a vital clinical variable. Physiological control systems that work based on the measuring of prelaod can improve perfusion, HF patients' quality of lives, and their lifespans, prevent incidence of the hazardous events such suction and congestion and increase the exercise capacity by providing higher pump flow [18], [19], [21]. Therefore, the proposed preload-based physiological controller can have the same benefits.

A physiological control for LVAD using MFAC was proposed in [19], [21] to maintain the preload in the physiological range of 3 to 15 mmHg for 100 different patient conditions. In the proposed control approach, a sensor for measuring the left ventricular pressure (LVP) was used and then the preload was extracted by a novel algorithm from the LVP in real-time mode. However, there is no commercial long-term implantable sensor for LVP measurement, making our solution that uses an accurate preload estimator more viable. In addition, the LVAD speed was controlled using the compact linearization form of MFAC, which is not able to identify all the plant dynamics. In contrast, in this study, the preload was estimated using deep CNNs based on the pump flow and FFDL-MFAC was used to identify the full dynamics of the CVS and LVAD and control the pump speed. Moreover, the use of an FFDL-MFAC controller in this study is advantageous because it does not have the common issues of other adaptive controllers such as ANN [17] and fuzzy controllers [20]. These controllers generate a high computational burden and rely on accurate rules for different patient conditions, respectively. Other controllers such as model predictive control (MPC) and optimal control need a simplified state-space model that cannot be generalized and modified for different patient conditions [45].

Frank Starling-like controllers have been proposed, which can restore physiological response of the native heart, can prevent suction, congestion and improve the patient's mobility [13]–[15], [18]. However, in these types of controller, clinicians need to manually set a desired flow across different patient conditions such as sleeping, walking or doing exercise. Furthermore, FS curves are patient-dependent. Finding the ideal FS curve for each patient requires frequent clinical check-ups, which makes implantation of Starling-like controller for different patients impossible. In contrast, a constant preload-based control strategy is simpler: the accurate preload estimation means that the clinician could adjust target preload without performing a speed ramp under the guidance of ultrasound. This is the reason that a preload-based physiological control system for LVADs was employed in this study, which can also prevent ventricular suction, pulmonary congestion and improve cardiac perfusion.

CNN can be implanted for application of super-resolution 4K ultra high-definition videos at 60 frames per second using FPGA in real-time mode [46]. Our proposed deep CNN is much simpler compared to the one proposed by Kim *et al.* [46]. FFDL-MFAC also needs very low computational burden since it is made by some simple equations. Therefore, there is no any issue for implementation of CNN and FFDL-MFAC with current technology in real-time.

This study has some limitations. First, the baroreflex effect was not considered in this study which may influence CVS hemodynamics and therefore control performance [47]. However, the baroreflex can be simulated as variations in CVS, which we have shown that the FFDL-MFAC controller can compensate for any changes in CVS well due to its adaptive structure.

Second, the proposed preload estimator in this study works based on the ideal pump flow estimator. Although there are no perfect pump estimators, several studies reported an estimator for pump flow with a very low value error between 0.002 to 0.38 L/min [48]–[50]. Further investigation should be done to improve the accuracy of pump flow estimators based on the pump intrinsic variables.

Third, although most physiological control systems for LVADs are evaluated through in-vitro studies, this study was performed only using numerical models of the CVS and LVAD. Ideally, the estimator would be trained on human or animal data. However, it is very time-consuming and expensive to obtain enough invasive pressure and flow data to robustly train a CNN. The advantage of using the numerical model (which was validated from animal experiments) was that it enabled us to easily simulate 2400 different patient conditions for validation of preload estimator. This would be a laborious and expensive process to complete in-vivo or in humans. Similarly, the numerical model facilitated simulation of 180 different patient conditions for validation of the preload estimation and physiological controller.

Future work will include validating the proposed estimator and the physiological controller via in-vitro and in-vivo experiments. Future work will also involve long term simulation of the proposed sensorless control system on patients with LVAD over the years via analysis of rapid changes of the CVS parameters. If the proposed method can handle these rapid changes as the worst-case scenarios, then it should be able to handle the long-term changes.

## V. CONCLUSION

In this study, a novel real-time deep CNN model for estimation of preload based on the LVAD flow was proposed. Furthermore, this estimator was combined with a new adaptive sensorless physiological control system for adjusting HVAD pump speed. This combined system was shown to be robust to a wide variety of inter- and intrapatient variations. When tested on 30 unseen patient datasets, no ventricular suction or pulmonary congestion events occurred. This shows that the inclusion of a preload estimator instead of an implantable pressure sensor that is likely to have long-term reliability issues, has no detrimental effects on the performance of the physiological controller.

## ACKNOWLEDGMENT

The authors would like to recognize the financial assistance provided by the National Health and Medical Research Council Centres for Research Excellence (APP1079421).